\documentstyle[prl,epsfig,aps,multicol]{revtex}

\title{Loss of Causality in Discretized Light-Cone Quantisation}

\author{T.~Heinzl$^{a}$, H.~Kr\"{o}ger$^{b}$ and N.~Scheu$^{c}$}

\address{$^{a}$ Theoretisch-Physikalisches Institut, 
  Friedrich-Schiller-Universit\"{a}t Jena, D-07743 Jena, Germany \\
  $^{b}$D\'{e}partement de Physique, Universit\'{e} Laval, Qu\'{e}bec,
  Qu\'{e}bec G1K 7P4, Canada \\ 
  $^{c}$Institut f\"{u}r Theoretische Physik, Universit\"{a}t Linz,
  A-4040 Linz, Austria }
      
\date{\today} 

\begin{document} 

\newcommand{\co}[1]{{\bf[#1]}}
\newcommand{\be}{\begin{equation}}
\newcommand{\ee}{\end{equation}}
\newcommand{\sgn}{\mbox{sgn}}
\newcommand{\slesssim}{{\scriptstyle \lesssim}}

\renewcommand{\vec}[1]{{\bf #1}}

\maketitle

\begin{abstract}
    We demonstrate that front form quantisation with periodicity in a
    compact light-like direction (``discretized light-cone quantisation'')
    violates microcausality. 
\end{abstract}

\pacs{Valid PACS appear here}

\begin{multicols}{2}

\section*{Introduction}
{\em Microcausality} is a fundamental postulate in relativistic
quantum field theory.  Mathematically, it states that two local
operators ${\cal O}_1(x)$ and ${\cal O}_2(y)$ at space-time positions
$x$ and $y$ must commute if $x-y$ is a space-like distance.
Physically, it means that signals cannot be transmitted faster than
with the velocity of light, $c$.  Imposing the requirement of
microcausality eliminates a large class of possible quantisation
schemes.  For instance, quantising scalar fields with 
anti-commutators violates microcausality, enforcing quantisation
of scalars in terms of commutators. Along these lines, the
celebrated {\em spin-statistics theorem} \cite{streater:63} is
established.

In this letter we demonstrate that the popular method of {\sl front
form (FF)} or {\sl light-cone (LC)} quantisation \cite{dirac:49} leads to
a breakdown of microcausality when space-time is compactified to a
cylinder with the periodic direction being chosen as light-like. As
the momenta conjugate to this direction become discrete, the method is
usually referred to as {\sl discretized light-cone quantisation}
(DLCQ) \cite{maskawa:76,brodsky:85}. Its range of applicability has
recently been extended to include M-theory \cite{susskind:97}.  

We will be concerned with a massive scalar field in $d$ space time
dimensions.  After some general remarks we will actually specialise to
$d$ =2. Our notations and conventions are
\begin{eqnarray}
\label{NOT}
  && x^{\pm} = x^{0} \pm x^{d-1} \; , \quad k^{\pm} = k^{0} \pm k^{d-1}
  \; , \\
  && x \cdot k = g_{\mu \nu} x^\mu k^\nu \; , \quad g_{+-} = g_{-+} =
  1/2  \; .
\end{eqnarray}
FF quantisation (for a recent review see \cite{heinzl:98}) amounts to
prescribing field commutators on the quantisation surface
$x^+=0$. This is a hyperplane tangent to the LC which in $d=2$
collapses to a light ray.

It has been shown by a number of authors
\cite{neville:71,steinhardt:80,heinzl:94}, that quantisation
on just one light-like surface is ambiguous. Knowledge of initial
conditions on two quantisation surfaces, say $x^{+}=0$ and $x^{-}=0$,
is necessary in order to have a well-posed (characteristic)
initial-value problem~\cite{heinzl:94}. As a result, the
characteristic initial values uniquely determine solutions to the
Klein-Gordon equation for all $x^\pm$ larger than the initial ones.

In 1994, however, Heinzl and Werner\cite{heinzl:94} were able to show
that the introduction of {\em periodic boundary conditions} (pBC) in
$x^-$ direction (assumed to be compact) uniquely determines the fields
on the second quantisation surface, $x^-=0$, in terms of the fields on
the first quantisation surface, $x^+=0$. An infinite-volume
formulation of this problem, however, is rather involved and in
general will require the use of distribution theory
\cite{grange:99,dimock:98}. It will not be addressed in this
letter. We will rather restrict to the finite-volume case and show
that prescribing light-like BC---though solving the initial-value
problem---is in conflict with microcausality. The results presented
here are built upon the Ph.D. thesis \cite{scheu:97} where the loss of
causality in DLCQ has been reported for the first time. Recently, this
finding has been confirmed by other researchers \cite{grange:99}.

\section*{Violation of Microcausality}
\noindent {\bf A. Generalities} \\
We start with a free scalar field $\phi$ in $d$ space-time dimension.
The commutator of two free scalar fields is of course known for all
times (i.e.~everywhere in Minkowski space), $[\phi(x),\phi(0)] = i
\Delta(x)$.  $\Delta(x)$ denotes the  Pauli-Jordan \cite{jordan:28} or
Schwinger \cite{schwinger:48} function, 
\begin{equation}
\label{DELTA_COV}
  \Delta(x) = \frac{1}{i} \int \frac{d^{d} k}{(2\pi)^{d-1}} ~
  \delta(k^{2} - m^{2}) ~ \text{sgn}(k^{0}) ~ e^{-i k \cdot x} .
\end{equation}
As the sign of $k^0$ does not change under proper orthochronous
Lorentz transformations, $\Lambda \in {\cal L}_+^\uparrow$, if $k^2$ is
spacelike, $\Delta$ is a Lorentz invariant function. The sign
function in addition guarantees that $\Delta$ is antisymmetric,
$\Delta(x) = \Delta (-x)$, as is necessary for a commutator.

It is well known that $\Delta(x)$ obeys microcausality. This is the
statement that $\Delta (x)$ has to vanish for $x$ space-like,
i.e.~$x^2 < 0$.  A very elegant argument to see this is due to
Gasiorowicz \cite{gasiorowicz:66}. For $x^2 < 0$, there is a Lorentz
transformation $\Lambda \in {\cal L}_+^\uparrow$ that takes $x$ to
$-x$, thus, by invariance, $\Delta (x) = \Delta (-x)$, for $x^2 <
0$. Therefore, outside the LC, $\Delta$ is both symmetric and
antisymmetric in $x$ and must vanish. The argument does not work for
$d=2$, as the regions $x^1 > 0$ and $x^1 < 0$ are
disconnected. Nevertheless, causality also holds in $d=2$, as one can
see upon evaluating (\ref{DELTA_COV}) for this case,
\begin{eqnarray}
\label{DELTA_2}
  \Delta (x) &=& - \frac{1}{2} \text{sgn} (x^0) \theta(x^2) J_0 (m
  \sqrt{x^2}) \; , \nonumber \\ 
  &=& - \frac{1}{4} \big[ \text{sgn}
  (x^+) + \text{sgn} (x^-) \big] J_0 (m \sqrt{x^2}) \; ,
\end{eqnarray}
which indeed vanishes outside the LC (here $J_0$ denotes the Bessel
function). The restrictions of (\ref{DELTA_2}) to $x^0 = 0$ and $x^+ =
0$ yield the canonical commutators of the two quantisation schemes
\cite{heinzl:98}.

In (\ref{DELTA_2}) we have given both the IF and FF versions of $\Delta$
which are, of course, simply related by the coordinate transformation
(\ref{NOT}). These two forms can actually be represented as one
dimensional integrals by performing the energy integrations over $k^0$
and $k^-$, respectively,
\begin{eqnarray}
  \text{IF:} \quad \Delta (x) &=&  - \int \frac{dk^1}{2 \pi \omega_k}
  \sin (\omega_k x^0 - k^1 x^1) \; , \label{DELTA_COV_IF} \;  \\
  \text{FF:} \quad \Delta (x) &=&  - \int_0^\infty \frac{dk^+}{2 \pi k^+}
  \sin(\hat k^- x^+ /2 + k^+ x^- /2) \; . \label{DELTA_COV_FF}
\end{eqnarray}
The on-shell values of the energies are given by $\omega_k = (k_1^2 +
m^2)^{1/2}$ and $\hat k^- = m^2/ k^+$. Note the restriction of the 
integration in
(\ref{DELTA_COV_FF}) which is due to the positivity of the
longitudinal momentum $k^+$. Both representations (\ref{DELTA_COV_IF})
and (\ref{DELTA_COV_FF}) can be integrated and yield
(\ref{DELTA_2}). As a cross check we note that (\ref{DELTA_COV_IF}) and
(\ref{DELTA_COV_FF}) are still related by the coordinate
transformation (\ref{NOT}) applied to the \emph{on-shell} momenta,
\begin{equation}
\label{BLM_TRANSF}
  k^\pm = \omega_k \pm k^1 = \sqrt{k_1^2 + m^2} \pm k^1 \; .
\end{equation}
This makes the positivity of $k^+$ explicit and entails that the
integration measures are related by the singular transformation
$dk^1 / \omega_k = dk^+ / k^+$.
Let us now investigate how (\ref{DELTA_COV_IF}) and
(\ref{DELTA_COV_FF}) get modified in a finite volume. To this end we
restrict the spatial coordinates, $ -L \le x^1, x^- \le L$, and impose
pBC for the field $\phi$. The conjugate momenta become discrete,
$k^1_n \equiv \pi n /L$ and $k_n^+ \equiv 2 \pi n /L$, respectively.
The finite volume representations are defined by replacing the
integrals (\ref{DELTA_COV_IF}) and (\ref{DELTA_COV_FF}) by the
discrete sums,
\begin{eqnarray}
  \Delta_{IF} (x) &\equiv& - \sum_{n = -N}^{N}
  \frac{1}{2 \omega_n L} \sin(\omega_n x^0 - n \pi x^1 /L) \; ,
  \label{DELTA_IF} \\ 
  \Delta_{FF} (x) &\equiv&
  - \sum_{n=1}^{N} \frac{1}{2 \pi n} \sin (\hat k^-_n x^+ /2 + n \pi x^- /
  L) \; , \label{DELTA_FF}
\end{eqnarray} 
where the limit $N \to \infty$ is understood. 
The on-shell energies for discrete momenta are defined as $\omega_n =
(n^2 \pi^2/L^2 + m^2)^{1/2}$ and $\hat k_n^- = m^2 L/2\pi n$. For both
functions, $\Delta_{IF}$ and $\Delta_{FF}$, the periodicity in $x^1$
and $x^-$, respectively, with periodicity length $2L$, is
obvious. Note that $\Delta_{IF}$ contains a zero mode ($n$ = 0), while
$\Delta_{FF}$ does not. This is a consequence of the Klein-Gordon
equation \cite{heinzl:94}.
In what follows we will show that, unlike $\Delta_{IF}$, $\Delta_{FF}$
does not obey microcausality, i.e.~does not vanish for $x^2 \equiv x^+
x^- < 0$. In addition, we find that $\Delta_{FF}$ does not converge to
$\Delta$ in the infinite volume limit. It actually turns out  that BC in
a light-like direction are quite generally incompatible with
causality.
\\

\noindent {\bf B. Numerical Results} \\
In general, the sums (\ref{DELTA_IF}) and (\ref{DELTA_FF}) cannot be
evaluated analytically.  Therefore we have calculated them
numerically. The summation cutoff $N$ has been chosen sufficiently
large to establish numerical convergence. The results are shown in
Fig.~1 and Fig.~2.

\begin{figure}
\begin{center}
\epsfig{figure=Fig1.epsf,height=5cm,width=.8\linewidth,angle=0}
\end{center}
\noindent
Fig.~1: $\Delta_{IF}(X,T)$ as a function of $X= x^1/2L$, for $T= x^0/2L =
0.2$, $mL = 1$, $N = 50$.
\end{figure}

\begin{figure}
\begin{center}
\epsfig{figure=Fig2.epsf,height=5cm,width=.8\linewidth,angle=0}
\end{center}
\noindent
Fig.~2: $\Delta_{FF}(v,w)$ as a function of $v = x^- /2L$, for $w= m^2Lx^+
/2 = 10000$, $N = 70$.
\end{figure}

\noindent By comparing the two figures one observes a striking
difference. $\Delta_{IF}$ is a smooth and regular function, while
$\Delta_{FF}$ looks `noisy' and irregular. Furthermore, for fixed $0 <
x^0 < L$, $\Delta_{IF}$ has compact support \emph{inside} the LC,
$-x^0 < x^1 < x^0$ (and in the periodic copies of this
interval). Outside the LC $\Delta_{IF}$ shows tiny oscillations around the value zero, which vanish in the limit $N \to \infty$. The oscillations are due to
Gibbs' phenomenon (the Fourier series does not converge uniformly in the vicinity of points where the limiting function makes jumps). 
Physically, what happens is that we have point sources
located at positions $x^1 = 2Ln$. These `emit' spherical waves which
do not interfere unless $x^0\ge L$. For $x^0 > L$ (not shown here) we have an interference phenomenon so that $\Delta_{IF}$ no longer vanishes outside the
LC, which is a straightforward consequence of periodicity.

The situation concerning $\Delta_{FF}$ is different. Numerically, one
sees that despite the irregular shape the sum (\ref{DELTA_FF})
converges to a periodic function. The most important observation,
however, is that $\Delta_{FF}$ does not vanish outside the LC,
i.e. for $x^- < 0$, $x^+ > 0$ as in Fig.~2. This a clear
\emph{violation of microcausality}. We have numerical evidence for a
corresponding behavior in $d=3,4$ space-time dimensions.
\\

\noindent {\bf C. Analytical Results} \\
Let us try to get an analytical understanding of the numerical results
beginning with $\Delta_{IF}$. A straightforward application of the
Poisson resummation formula yields
\be
\label{DELTA_IF_PER}
  \Delta_{IF} (x) = \sum_n \Delta (x^0, x^1 + 2Ln) \; ,
\ee
with the continuum $\Delta$ from (\ref{DELTA_2}). This result is
exactly what we see in Fig.~1, a periodic array of (nearly) smooth
functions with support inside the LC (and its periodic copies). 
It should be stressed that $\Delta_{IF}$ is causal even
for \emph{finite} extension $L$, i.e.~without the infinite-volume limit
being performed.  

Let us now analyze $\Delta_{FF}$. First note the rather weak
localization properties of $\Delta$ in the LC direction $x^-$. For
positive LC time $x^+$, $\Delta$ vanishes outside the LC, i.e. for
$x^- < 0$, and decays slowly for $x^- > 0$, asymptotically like
$(x^-)^{-1/4}$. The integrand in (\ref{DELTA_COV_FF}), denoted by
$I(k^+)$, oscillates rapidly for small $k^+$ such that the zero mode,
i.e.~the limit $I(k^+ = 0)$, is not defined. It turns out that
this makes the application of Poisson resummation impossible.  Because
the latter cannot be used, let us consider the following
alternative which leads to an analytic and close
approximation of $\Delta_{FF}$. For this purpose we rewrite
(\ref{DELTA_FF}) in terms of dimensionless variables (cf.~Fig.~2), 
\be
\label{DIMLESS}
  \Delta_{FF} (v,w) = \sum_{n=1}^N \frac{1}{2\pi n} \sin(w/2 \pi n + 2 \pi n
  v) \; , 
\ee
with $v \equiv x^- / 2L$, $w \equiv m^2 L^2 (x^+ / 2L)$. If we expand
(\ref{DIMLESS}) in powers of $w$, the sum over $n$ can actually be
performed using summation formulae 1.443.1/2 from
\cite{gradshteyn:65}. With the restriction $-1 \le v \le 1$, the result is
\be
\label{BERNOULLI}
  \Delta_{FF} (v,w) = -\frac{1}{2} \sum_{n = 0}^N \frac{w^n}{n!
  (n+1)!} \sgn^{(n+1)} (-v) B_{n+1} (|v|) \; ,
\ee
where $B_n$ denotes the $n$th Bernoulli polynomial
\cite{gradshteyn:65}.  The series (\ref{BERNOULLI}), being a
power series instead of a Fourier series, converges rather rapidly as
a function of $w$. In addition, the limit is approached uniformly (no
Gibbs phenomenon). This is obvious from Fig.~3 where we compare the
resummed expression (\ref{BERNOULLI}) with the Fourier representation
(\ref{DIMLESS}) (for $w$ = 5). The agreement is quite impressive
which is to be expected as we have summed the first 20 terms in
(\ref{BERNOULLI}).

\begin{figure}
\begin{center}
\epsfig{figure=Fig3.epsf,height=5cm,width=.8\linewidth,angle=0}
\end{center}
\noindent
Fig.~3: Comparison of the Fourier representation
(\protect\ref{DIMLESS}) (for $N = 40$) with the result of Bernoulli
resummation (\protect\ref{BERNOULLI}), for $w = 5$, $-1 \le v \le 1$,
$N = 20$.
\end{figure}

Numerically, one finds convergence if the number $N$ of terms summed
over is of the order of $w/2$. This is due to the fact that the
amplitude of the Bernoulli polynomials $B_n$ decreases rapidly with
$n$. Thus, for $w \slesssim 2$, i.e.~$x^+ \slesssim 4/m^2 L$, the
first two terms in the expansion (\ref{BERNOULLI}) are already a
rather good approximation so that we can write,
\be
\label{APPROX}
  \Delta_{FF} (v,w) \simeq -\frac{1}{4} \sgn(v) + \frac{v}{2} -
  \frac{w}{4} (v^2 - |v| + 1/6)  \; .
\ee
This result provides an \emph{analytical} check that $\Delta_{FF}$
does not vanish outside the LC ($-1 < v < 0$), and thus, that
causality is violated.
\\

\section*{Restoration of a Causal Propagator}
With representation (\ref{DELTA_FF}) we are sampling
a continuous function $I(k^+)$ by \emph{equidistant} points on a
momentum grid in $k^+$. For small $k^+$, however, this is a very bad
approximation, as $I(k^+)$ is rapidly oscillating there, with a
frequency increasing roughly as $1/k^+$. The point $k^+ = 0$ is thus
an accumulation point of the Fourier spectrum. In its vicinity, we
should actually sample with a momentum resolution $\triangle k_n^+
\sim 1/n$. In other words, instead of harmonic one should use
\emph{anharmonic} ``Fourier'' analysis. If we use (\ref{BLM_TRANSF}) to introduce
\emph{new} discrete longitudinal momenta,
\begin{equation}
  k_n^\pm \equiv \omega_n \pm k_n^1 = \frac{1}{L} (\sqrt{n^2 \pi^2 + m^2
  L^2} \pm n \pi) \; ,
\label{NEWGRID}
\end{equation}
we can write down the causal, finite-volume commutator in
terms of light-cone variables,
\begin{equation}
\label{DELTA_C}
  \Delta_c (x) = - \frac{1}{L} \sum_n \frac{k_n^+}{(k_n^+)^2 + m^2}
\sin (k_n \cdot x) \; ,
\end{equation}
where $k_n \cdot x = k_n^- x^+ / 2 + k_n^+ x^- / 2$.  Obviously, the
momentum grid (\ref{NEWGRID}) used in (\ref{DELTA_C}) is not
equidistant. In particular, for small $k_n^+$, corresponding to large
negative $k_n^1$, one finds $\triangle k_n^+ \sim 1/n$. Thus, the
small-$k^+$ region becomes sampled in a reasonable way such that the
features of $I(k^+)$, which guarantee the causality of $\Delta$ are
properly described even on a finite (momentum) lattice (see Fig.~4).

\begin{figure}
\begin{center}
\epsfig{figure=Fig4.epsf,height=5cm,width=.8\linewidth,angle=0}
\end{center}
\noindent
Fig.~4: The causal commutator (\protect\ref{DELTA_C}) in the LC
representation as a function of $v = x^- / 2L$, for $x^+ / 2L = 0.2$, $mL
= 50$, $N = 50$. It vanishes for $-1 < v < 0$ (up to the unavoidable
Gibbs phenomenon).
\end{figure}

\section*{Discussion}
The above analysis shows that one {\em cannot} have both, periodicity
in $x^-$ and causality of the commutator $\Delta(x^+ , x^-)$. If one
insists on periodicity, one violates causality and vice
versa. Consequently, the method of DLCQ in which the field operators
are expanded in periodic plane waves, is non-causal. This does not
come as too big a surprise: it is well known that DLCQ yields a rather
poor representation of the small-$k^+$ behaviour of observables
\cite{vandeSande:96}. Using the relation between Pauli-Jordan function
and  Feynman propagator, $\pi \Delta (k) = - \sgn (k^+) \mbox{Im}
\Delta_F (k)$, we see that a causality violation also affects the Feynman propagator
$\Delta_F$. Note that the causality of $\Delta$ can be viewed as a
delicate cancellation between particle and anti-particle propagation
amplitudes. Therefore it seems that also charge conjugation symmetry
is violated by imposing light-like periodicity. Furthermore, with
microcausality being at the heart of any dispersion relation, one
expects problems also there. We have seen that one can taylor an {\em
ad hoc} momentum grid with a special anharmonic resolution which
remedies the causality violation of the \emph{commutator}. It is an
open question, however, whether this solves the causality problem of
DLCQ in general. We expect the answer to be negative: any causal Green
function will have its own peculiar small-$k^+$-behaviour and thus
will require its own momentum grid which generically will be different
from the one introduced above. An ensuing dependence of results on a
particular discretisation choice clearly cannot be accepted. \\

\noindent {\bf Acknowledgements} \\
T.H.~thanks E.~Elizalde, F.~Lenz, B.~Schroer, T.~Tok, and A.~Wipf for
valuable hints and discussions. The authors gratefully acknowledge
support by DFG, project WI 777/3-2 (T.H.), NSERC Canada (H.K.) and
Austrian Science Fund, project P111098 PHY (N.S.).

\end{multicols}






\end{document}